\documentclass[fleqn,usenatbib]{mnras}

\usepackage{newtxtext,newtxmath}

\usepackage[T1]{fontenc}

\DeclareRobustCommand{\VAN}[3]{#2}
\let\VANthebibliography\thebibliography
\def\thebibliography{\DeclareRobustCommand{\VAN}[3]{##3}\VANthebibliography}

\usepackage{graphicx}	
\usepackage{amsmath}	
\usepackage{hhline}     
\usepackage{subcaption} 
\usepackage{xcolor}
\usepackage{array} 

\defcitealias{Li_Youdin_2021}{LY21}
\newcolumntype{R}{>{\color{red}}l}

\title[Dust trapping]{Dust trapping in protoplanetary discs after stellar flybys}

\author[V. R. Prasad et al.]{
Vasundhara R. Prasad,$^{1}$\thanks{E-mail: vprasad@ast.cam.ac.uk}
Cristiano Longarini,$^{1}$
and Cathie J. Clarke$^{1}$
\\
$^{1}$Institute of Astronomy, University of Cambridge, Madingley Road, Cambridge CB3 0HA, United Kingdom
}

\date{Accepted XXX. Received YYY; in original form ZZZ}

\pubyear{\the\year{}}

\begin{document}
\label{firstpage}
\pagerange{\pageref{firstpage}--\pageref{lastpage}}
\maketitle

\begin{abstract}
Stellar flybys are likely to be common in young star-forming regions and could be responsible for substructures observed in protoplanetary discs. Using three-dimensional smoothed particle hydrodynamics simulations, we study dust trapping in discs perturbed by parabolic coplanar flybys. We find that spiral structures are induced in the gas and dust discs for both prograde and retrograde encounters. By tracking individual dust particles within the flyby-induced substructures, we determine that they have a highly enhanced dust to gas ratio compared to particles in an unperturbed disc. We further find that the local dust to gas ratios in flyby-induced substructures are sufficiently high to trigger the streaming instability and hence facilitate planetesimal formation in young discs.
\end{abstract}




\section{Introduction}

An open question in planet formation is the growth of dust grains in young protoplanetary discs. Micrometre-sized dust grains can grow via collisional sticking, but at centimetre/metre sizes dust growth is thought to be hindered by the fragmentation barrier \citep{Blum_Wurm_2008,Birnstiel_2024} and by rapid radial drift of dust grains into the star as they become decoupled from the gas \citep{Weidenschilling_1977,Takeuchi_Lin_2002}. 

One mechanism proposed to overcome this barrier is the streaming instability (SI; \citealt{Youdin_Goodman_2005}). The SI arises from a back-reaction of dust grains on gas that cause the grains to converge into azimuthal filaments, which gravitationally collapse when they reach the critical Roche density and trigger planetesimal formation. However, the SI is only triggered at sufficiently large values of the local surface dust-to-gas ratio (\citealt*{Carrera_2015}; \citealt*{Yang_2017}; \citealt{Li_Youdin_2021}, hereafter \citetalias{Li_Youdin_2021}). It therefore appears that the growth of dust grains to planetesimal scales requires the presence of dust overdensities in the protoplanetary disc.

Recent observations by the Atacama Large Millimeter/submillimeter Array (ALMA) have revealed that protoplanetary discs near-ubiquitously have substructures such as gaps, rings, cavities and spirals \citep{Andrews_2020}. It has been proposed that the pressure maxima of these discs may act as ``dust traps'' and concentrate dust particles in overdense regions \citep*{Birnstiel_Fang_Johansen_2016}. Possible mechanisms for the formation of these dust traps include gaps opened by planets (\citealt*{Pinilla_2012}, \citealt{Ayliffe_Laibe_Price_Bate_2012}), spirals induced by gravitational instabilities \citep{Dipierro_2015} or self-induced dust traps \citep{Gonzalez_2017}. These dust traps could be key to triggering the SI and the formation of planetesimals from small dust grains. For this reason, substructures in young discs are widely viewed as potential birthplaces of protoplanets.

One possible mechanism for perturbing a young protoplanetary disc is via encounters with other stars. Findings by \citet{Pfalzner_2013} imply that close to 50\% of young solar-type stars located within $0.5 \text{ pc}$ of a cluster centre will undergo an encounter within $1000 \text{ AU}$; \citet{Pfalzner_Govind_2021} further argued that close-in flybys are more common than previously thought in low-mass stellar clusters. \citet*{Cuello_Ménard_Price_2023} also argue that at least 75\% of young stars are born in regions of high stellar density where encounters with other stars are common, suggesting that stellar encounters could play a significant role in shaping the dynamics of young protoplanetary discs.

We here define a \textit{stellar flyby} as a non-recurring event wherein a star on an unbound parabolic or hyperbolic orbit perturbs another star. \citet{Clarke_Pringle_1993} found that flybys can lead to tidal stripping, the formation of spiral arms within the disc, disc truncation and the capture by the perturber of some disc material. Several of these processes could locally or globally enhance the dust to gas ratio in the disc, potentially triggering the SI and hence planetesimal formation. We have observed several systems that are candidates for ongoing or recent flybys, such as RW Aur \citep{Cabrit_2006,Dai_2015,Rodriguez_2018}, HV Tau and DO Tau \citep*{Winter_Booth_Clarke_2018}, UX Tau \citep{Menard_2020} and Z CMa \citep{Dong_2022}. There is therefore a strong motivation to investigate the ability of flyby-induced substructures to trap dust and facilitate planetesimal formation via the SI.

\citet{Cuello_2019} conducted a detailed study into the evolution of both gas and dust during and after a stellar flyby using three-dimensional smoothed particle hydrodynamic simulations. They found that the flyby induced substructures such as spirals and stellar bridges (material connecting the two stars) for both prograde and retrograde encounters, although the spiral structures were sharper and less radially extended in the dust than in the gas, showing that gas and dust respond differently to flyby-induced perturbations. They also found that flybys accelerate the radial drift of dust towards pressure maxima in the disc, potentially triggering the streaming instability or dust trapping, and that flybys result in a global increase of dust to gas ratio in the disc. They suggest that spirals can act as dust traps, mixing dust grains of different sizes, but do not investigate this in detail. 

\citet{Smallwood_2023} also conducted 3D smoothed particle hydrodynamic simulations of gaseous protoplanetary discs perturbed by stellar- and sub-stellar mass perturbers, and studied the formation and long-term evolution of the spiral arms produced by these flybys. The authors speculate that the spiral arms could act as dust traps and trigger planetesimal formation.

In this work we study the dust dynamics of protoplanetary discs perturbed by coplanar stellar flybys. We focus on the behaviour of the substructures induced in the gas and dust discs, with the aim of determining whether or not these substructures act as dust traps. We discuss the hydrodynamical code and simulation set-up in Section~\ref{sec:methods} and present their results in Section~\ref{sec:results}. Section~\ref{sec:discussion} discusses the implication of these results for planetesimal formation via the streaming instability. We summarise our findings and present our conclusions in Section~\ref{sec:conclusions}.

\section{Numerical Methods} \label{sec:methods}

We model protoplanetary discs perturbed by flyby encounters using the 3D smoothed particle hydrodynamics (SPH) code \textsc{Phantom} \citep{Price_2018}. SPH (for a review see \citealt{Price_2012}) is well-suited to simulating flyby encounters because there is no preferred geometry and angular momentum is accurately conserved. \textsc{Phantom} has been extensively tested for the case of parabolic flyby encounters \citep{Cuello_2019,Menard_2020,Smallwood_2023,Smallwood_2024,Nealon_2025} and also for the evolution of dusty discs \citep{2015MNRAS.453L..73D,Longarini_2021}. We simulated six coplanar prograde and retrograde flyby encounters using \textsc{Phantom}, varying the orbital parameters (namely the perturber mass $M_2$ and the periastron distance $r_\text{peri}$) of the flyby, and additionally one reference simulation  with no flyby encounter. A summary of the simulations is given in Table \ref{tab:simulations}.

\subsection{Initial conditions}

\subsubsection{Circumprimary disc setup}

We set up a coplanar (midplane at $z=0$) circumprimary disc orbiting a star of mass $M_1 = 1 M_{\sun}$ modelled as a sink particle. The disc has inner radius $R_\text{in} = 10 \text{ AU}$ and outer radius $R_\text{out} = 150 \text{ AU}$. The disc has a total gas mass of $0.01 M_{\sun}$ and is flared with an aspect ratio of $H/R = 0.05$ at $R=R_\text{in}$, increasing to $H/R = 0.1$ at $R=R_\text{out}$. The accretion radius for the primary star is set at $R_\text{acc}=10 \text{ AU}$. 

The equation of state of the disc is chosen to be locally isothermal (following \citet{Lodato_Pringle_2007}) and has a sound speed

\begin{equation}
    c_s \propto \left( \frac{R}{1 \text{ AU}} \right)^{-q},
\end{equation}

\noindent where $q=0.25$. The value of the sound speed is set by a locally isothermal disc temperature profile given by 

\begin{equation}
    T (R) = 64.0 \text{K} \left (\frac{R}{R_\text{in}} \right)^{-0.5}.
\end{equation}

The surface density profile of the disc is given by \citep{Lodato_Pringle_2007}

\begin{equation}
    \Sigma (R) = \Sigma_0 \left (\frac{R}{R_\text{in}} \right)^{-p} \left( 1 - \sqrt{\frac{R_\text{in}}{R}}\right),
    \label{eqn:surface_density}
\end{equation}

\noindent where $p=1$ and $\Sigma_0 = 17.1 \text{ g/cm}^2$ is a normalisation constant set by the disc's mass. The effects of disc self-gravity are ignored throughout this work. We adopt a mean \citet{Shakura_Sunyaev_1973} viscosity parameter $\alpha_\text{SS} = 0.005$ by setting a fixed ``artificial viscosity'' $\alpha_\text{AV} = 0.26$ and following the prescription in \citet{Lodato_Price_2010},

\begin{equation}
    \alpha_\text{SS} \approx \frac{\alpha_\text{AV}}{10}\frac{\langle h \rangle}{H}
\end{equation}

\noindent where $\langle h \rangle$ is the mean smoothing length.

We set up the disc with $10^6$ Lagrangian SPH gas particles and $10^5$ SPH dust particles. This number of SPH particles is justified following \citet{Smallwood_2023}, who performed a resolution study where the number of gas particles in the disc was increased from $10^6$ to $10^7$. The authors found that the lifetime of the spiral arms induced during the flyby was longer in the higher-resolution study by an order of less than 10 per cent, and that overall the differences between the two simulations were minimal. They considered that simulations with $10^6$ SPH particles sufficed to study the dynamics of the flyby-perturbed disc.

\subsubsection{Perturber setup} \label{sec:perturber_setup}

We place the perturbing star on a coplanar parabolic orbit. We restrict simulations to parabolic encounters ($e=1$) only because these encounters have the largest star-to-disc momentum transfer and so are likely to induce more prominent disc substructures \citep{Smallwood_2023}. Flybys on hyperbolic orbits are less efficient in capturing material \citep{Smallwood_2023} and are also physically unlikely in low-density stellar environments \citep{Winter_Booth_Clarke_2018}. The perturbing star arrives from the negative $y$-direction, leaves towards the negative $y$-direction, and is closest to the primary star (periastron) on the $y$-axis (i.e. the line $x=0$). The perturbing star's initial distance from the primary star is set to be 10 $r_\text{peri}$, and the encounter is evolved until the distance between the two stars is once again 10 $r_\text{peri}$ in all cases except the \texttt{retro} simulation, which was evolved for longer in order to study the long-term evolution of the disc. We set the accretion radius for the perturbing star at $5 \text{ AU}$.

We further restrict the flyby trajectories to solely coplanar orbits. Although purely coplanar orbits are naturally less physically likely than inclined flybys, \citet{Cuello_2019} and \citet{Smallwood_2023} note that prograde coplanar encounters generate more prominent spiral arms with higher peak surface densities than inclined prograde encounters. Hence coplanar encounters are most useful for a first study into whether or not dust trapping occurs in the spirals produced by flyby encounters. Furthermore, we simulated four prograde flybys and only two retrograde flybys, as we expect that prograde encounters will induce more prominent spiral arms than retrograde encounters \citep{Clarke_Pringle_1993,Cuello_2019}.

\begin{table}
\centering
\caption{Setup of the seven hydrodynamical simulations.}
\label{tab:simulations}
\begin{tabular}{cccc}
    \hline
    Simulation ID & $M_2$ [$M_{\sun}$] & $r_{\text{peri}}$ [au] & flyby inclination \\ \hhline{====}
    no\_flyby & 0.0 & $\infty$ & none \\ \hline
    standard\_run & 1.0 & 175.0 & prograde \\ \hline
    half & 0.5 & 175.0 & prograde \\ \hline
    intermediate & 1.0 & 262.5 & prograde \\ \hline
    far & 1.0 & 350.0 & prograde \\ \hline
    retro & 1.0 & 175.0 & retrograde \\ \hline
    int\_retro & 1.0 & 262.5 & retrograde \\ \hline
\end{tabular}
\end{table}

We choose periastron distances $r_\text{peri}$ ranging from 175 AU to 350 AU, which (for a disc extending to 150 AU) classify as either grazing or non-penetrating encounters. We do not simulate any encounters with smaller periastron distances, as we expect penetrating encounters to destroy the circumprimary disc instead of inducing substructures. We choose a perturber of equal mass to the primary star ($1 M_{\sun}$) in all encounters except the \texttt{half} simulation, where the perturber's mass is $0.5 M_{\sun}$.

\subsection{Dust properties}

We set up a dust disc with the same initial surface density profile (following Equation \ref{eqn:surface_density}) and radial extent as the gas disc, but with a mass of $10^{-4} M_{\sun}$ (i.e. an initial dust to gas ratio of 0.01). We choose an intrinsic dust grain density $\rho_\text{gr} = 3 \text{ g/cm}^3$ and a dust grain size $s=3 \text{ cm}$. In the Epstein drag regime the Stokes number, which describes the aerodynamical coupling of gas to dust, is given by \citep{epstein1924}

\begin{equation}
    \text{St} = t_s \Omega_k = \frac{\rho_\text{gr}s}{\rho c_s}\Omega_k,
    \label{eqn:stokes}
\end{equation}

\noindent where $t_s$ is the stopping time (the timescale on which dust grains couple to the flow of the surrounding gas), $\Omega_k$ is the Keplerian angular frequency, $\rho=\rho_g+\rho_d$ is the \emph{total} local density of gas and dust combined, and $c_s$ is the gas sound speed. Here we choose a dust grain size corresponding to $\text{St} \approx 10$.

We simulate $10^5$ dust particles using the ``two-fluid'' implementation in \textsc{Phantom} \citep{Laibe_Price_2012a,Laibe_Price_2012b}, following the finding by \citet{Cuello_2019} that for dust grain sizes larger than 1 mm
(i.e. Stokes numbers greater than 1) the two-fluid method is more appropriate than the one-fluid implementation (which models the dust and gas as a single type of particle). In these simulations, we are using the velocity reconstruction procedure presented in \citet{Price_2020}. For the \texttt{standard\_run} simulation, after 7340 yr (equivalent to 83\% of the flyby time as defined in \autoref{tab:timescales}), we use an implicit drag scheme for computational convenience, following \citet{implicit_drag1,implcit_drag2}.

\subsection{Dust tracking algorithm} \label{sec:dust_tracking_algorithm}

In order to determine whether or not dust trapping occurs in the flyby-induced substructures, we developed a simple algorithm to ``trace'' dust particles over the course of the simulation. Using a Lagrangian code here is extremely advantageous, as it allows us to isolate single SPH dust particles and follow their properties with time. 

We selected particles to track based on the presence within a gas or dust substructure at a given time using the following procedure to select 50 particles from the appropriate substructure:

\begin{enumerate}
    \item divide the disc into 50 equal azimuths;
    \item if choosing particles within \emph{gas} substructures, interpolate the gas density onto the dust particles using nearest-neighbour interpolation;
    \item exclude particles that are not gravitationally bound to the primary star at the end of the simulation (either because they were accreted or else captured by the perturber);
    \item further restrict the disc to only \emph{dust} particles within a specified radius range in order to select a specific substructure;
    \item within each azimuth, select the dust particle with the highest interpolated gas density (if tracing particles within gas substructures) or dust to gas ratio (if tracing particles within dust substructures);
    \item record the properties of the 50 dust particles chosen over the course of the entire simulation and take their mean values as a function of time.
\end{enumerate}

\section{Results} \label{sec:results}

\subsection{Global gas and dust evolution}

As discussed in \autoref{sec:perturber_setup}, we evolve each flyby encounter for the time it takes for the perturber's distance to the primary star to go from $10 r_\text{peri}$ to $r_\text{peri}$ and back to $10 r_\text{peri}$ in all cases except the \texttt{retro} simulation, which was evolved for longer. Because the timescale of the parabolic flyby varies depending on the periastron distance of the perturber orbit, when comparing the results of different simulations below we scale the flyby timescale so that $t=0.5$ when the perturber is at periastron ($t_\text{peri}$) in each case. The \texttt{no\_flyby} simulation, which cannot be scaled according to $t_\text{peri}$, uses the same scaling as the \texttt{standard\_run} and \texttt{retro} simulations. These timescales are summarised in \autoref{tab:timescales}. The dynamical timescale at the original outer edge of the disc at 150 AU is given by $\Omega_0^{-1}\approx1837$ yr, meaning that the simulations take place over between 5 and 13 dynamical timescales depending on the flyby's periastron distance.

\begin{table}
\centering
\caption{Timescales of the six flyby encounters. When comparing the time evolution of different simulations we set $2t_\text{peri}=1$. We also present the flyby timescale in units of the dynamical timescale at the original outer edge of the disc, $\Omega_0^{-1}$.}
\label{tab:timescales}
\begin{tabular}{ccc}
    \hline
    Simulation ID & $2t_\text{peri}$ [yr] & $2t_\text{peri}$ [$\Omega_0^{-1}$] \\ \hhline{===}
    standard\_run & 8843 & 4.8 \\ \hline
    half & 10212 & 5.6 \\ \hline
    intermediate & 16250 & 8.8 \\ \hline
    far & 25012 & 13.6 \\ \hline
    retro & 8843 & 4.8\\ \hline
    int\_retro & 16246 & 8.8\\ \hline
\end{tabular}
\end{table}

\begin{figure*}
    \centering
    \includegraphics{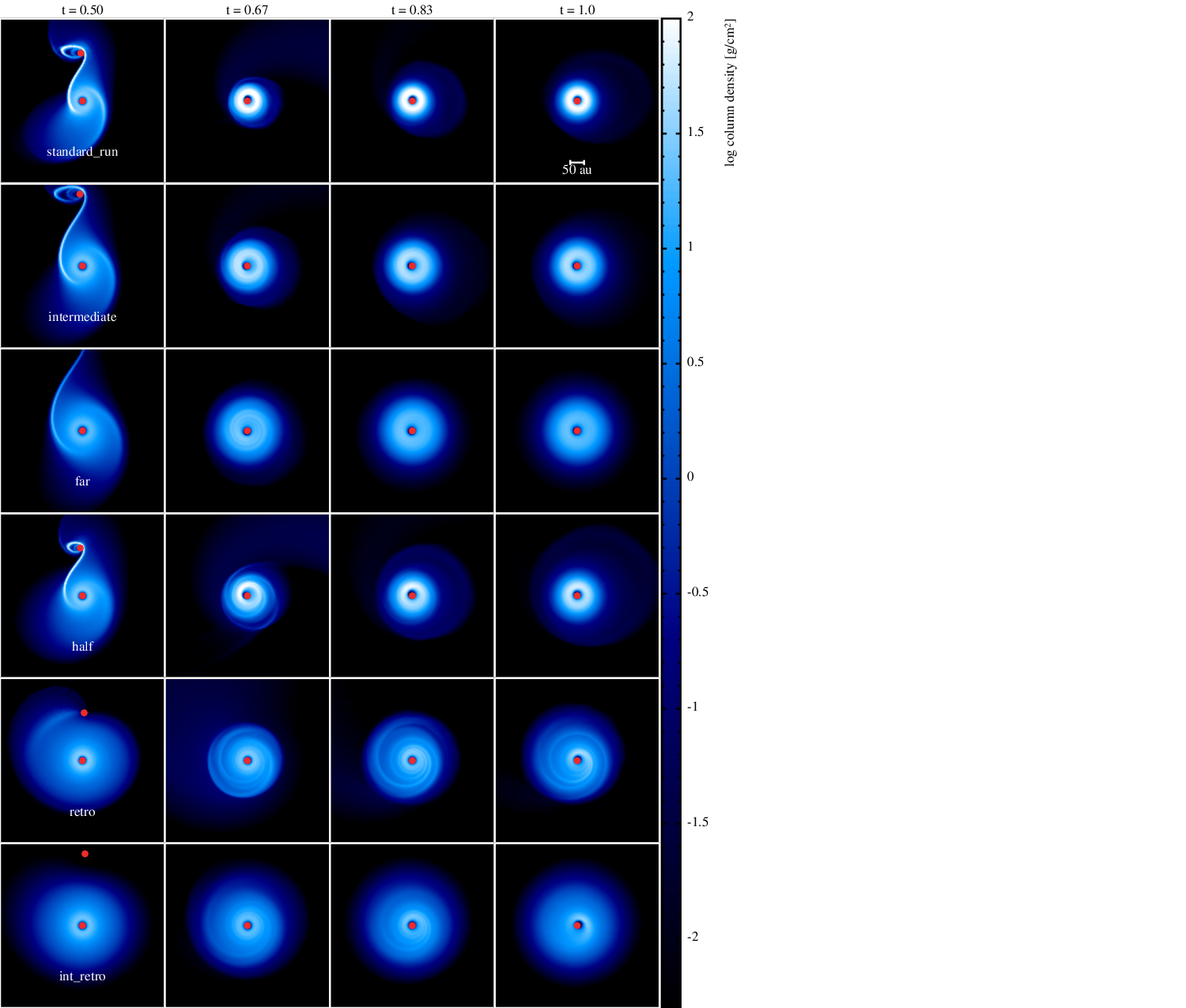}
    \caption{Face-on view of the gas column density of the primary disc for (top to bottom): the \texttt{standard\_run}, \texttt{intermediate}, \texttt{far}, \texttt{half}, \texttt{retro} and \texttt{int\_retro} simulations. Columns (left to right) are at times $t=0.5$ (periastron), $t=0.67, t=0.83$ and $t=1$. Sink particles (in red) are large for visualisation purposes only. In each of the prograde encounters we observe a bridge of material connecting the two stars, and (in some cases) the induction of a two-armed spiral structure which then dissipates. In the retrograde encounters the spiral structures persist for longer. }
    \label{fig:gas_snapshots}
\end{figure*}

\begin{figure*}
    \centering
    \includegraphics{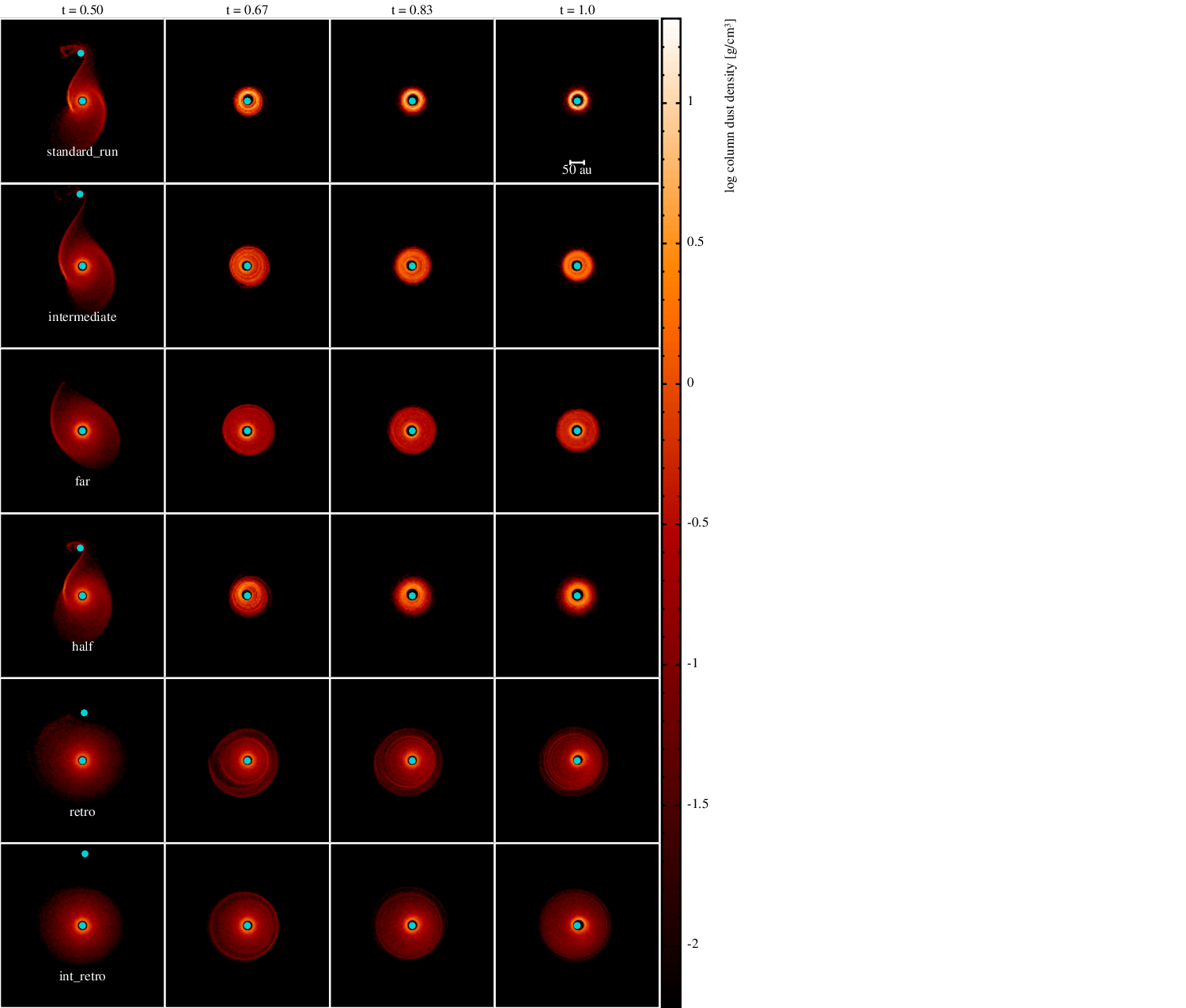}
    \caption{Face-on view of the dust column density of the primary disc for (top to bottom): the \texttt{standard\_run}, \texttt{intermediate}, \texttt{far}, \texttt{half}, \texttt{retro} and \texttt{int\_retro} simulations. Columns (left to right) are at times $t=0.5$ (periastron), $t=0.67, t=0.83$ and $t=1$. Sink particles (in blue) are large for visualisation purposes only. In the prograde encounters the dust disc is significantly truncated. Ring-like substructures of higher dust density appear when the periastron distance is larger than 175 AU, as well as in both the retrograde encounters.}
    \label{fig:dust_snapshots}
\end{figure*}

Figures~\ref{fig:gas_snapshots} and \ref{fig:dust_snapshots} show snapshots of the gas and dust column density at times $t=0.5$ (periastron), $t=0.67, t=0.83$ and $t=1$ for (top to bottom): the \texttt{standard\_run}, \texttt{intermediate}, \texttt{far}, \texttt{half}, \texttt{retro} and \texttt{int\_retro} simulations. In each of the four prograde simulations, we observe a two-armed spiral forming at periastron, with a bridge of material connecting the primary and perturber. At later times we see spiral structures in the gas density for the \texttt{intermediate}, \texttt{far} and \texttt{half} encounters, all of which have dissipated by $t=1$. In the two retrograde encounters, there is no bridge between the two stars but the two-armed spiral structure in the gas column density persists for longer. We observe a similar trend in the dust disc evolution shown in \autoref{fig:dust_snapshots}, where the flyby-induced substructures persist longer in the two retrograde cases than in the prograde encounters. The dust substructures are azimuthally symmetric rings rather than the spirals we see in the gas density. They are also considerably radially sharper (i.e. narrower in width) than the gas spirals. By $t=0.5$ a region of high dust density has formed at the inner edge of the disc as a result of radial drift of the dust \citep{Weidenschilling_1977}. We also observe very significant truncation of the dust disc in the prograde encounters, most notably the \texttt{standard\_run} simulation. 

\begin{figure}
    \centering
    \begin{subfigure}{\columnwidth}
        \includegraphics{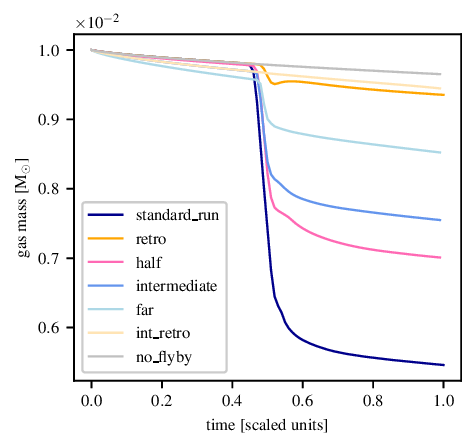}
        \caption{Mass of gas in primary disc over course of flyby.}
        \label{fig:gas_mass}
    \end{subfigure}
    \begin{subfigure}{\columnwidth}
        \includegraphics{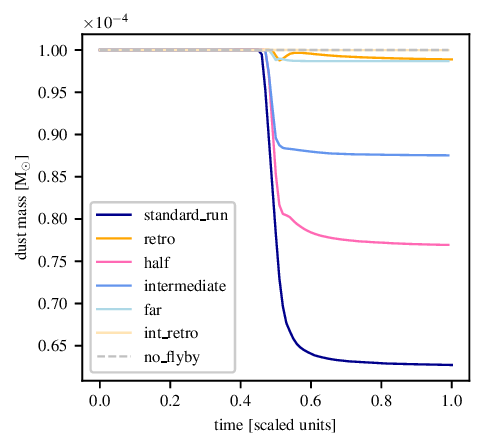}
        \caption{Mass of dust in primary disc over course of flyby. Note that the dust mass in both the \texttt{int\_retro} and \texttt{no\_flyby} simulations stay nearly constant at $10^{-4}M_{\sun}$.}
        \label{fig:dust_mass}
    \end{subfigure}
    \caption{Mass of the gas and dust components of the primary disc over the course of each flyby in scaled time units. Perturbers on prograde orbits capture significantly more material from the disc than retrograde perturbers.}
    \label{fig:mass_components}
\end{figure}

\begin{figure}
    \centering
    \includegraphics{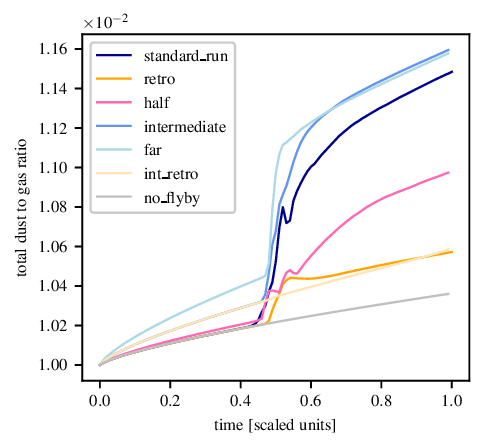}
    \caption{Total dust to gas ratio of the primary disc over the course of each flyby in scaled time units, defined as the ratio of the gas mass to the dust mass in the primary disc.}
    \label{fig:dtg_comparison}
\end{figure}

We further investigate the evolution of gas and dust by studying their total mass over time. \autoref{fig:mass_components} shows the mass of the gas and dust in the \emph{primary} disc over the course of the encounter. We observe significant mass loss from the disc after periastron in each of the prograde encounters, most notably the \texttt{standard\_run} case where over 40\% of the disc's mass is lost. The two retrograde encounters simulated are much less destructive with negligible mass stripped away from the primary disc by the perturber, and the very small decrease in disc mass in the \texttt{no\_flyby} simulation is explicable by accretion onto the central star. The mass loss is differential in all cases, with the perturber stripping away a greater proportion of the disc's gas than dust; this can be attributed to radial drift in the disc prior to the flyby, leaving the outer disc dust-poor. This leads to an increase in the total dust to gas ratio of the disc of up to 16\% in the prograde encounters, and a smaller increase of about 4\% in the retrograde encounters, as can be seen in \autoref{fig:dtg_comparison}.

\begin{figure}
    \centering
    \begin{subfigure}{\columnwidth}
        \includegraphics{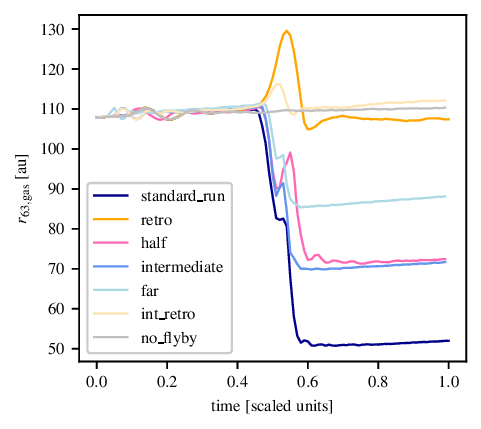}
        \caption{Radius of gas in primary disc over course of flyby.}
        \label{fig:gas_r63}
    \end{subfigure}
    \begin{subfigure}{\columnwidth}
        \includegraphics{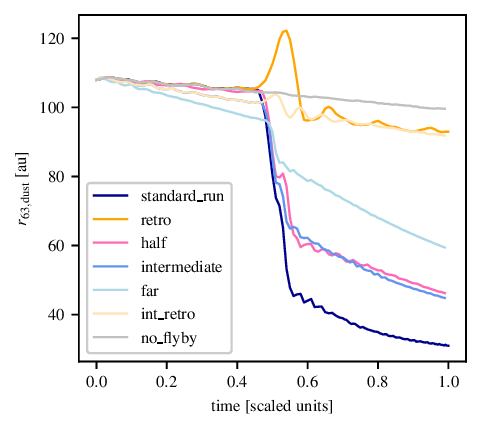}
        \caption{Radius of dust in primary disc over course of flyby.}
        \label{fig:dust_r63}
    \end{subfigure}
    \caption{Radius of the gas and dust discs over the flyby in scaled time units, calculated as the radius at which 63.2\% of the total (\subref{fig:gas_r63}) gas or (\subref{fig:dust_r63}) dust mass of the primary disc is enclosed.  Dust discs undergo more truncation than gas discs in each encounter. The large peak in radius at periastron for the \texttt{retro} simulation can be attributed to the disc ``puffing up'' as the perturber passes through it.}
    \label{fig:r63_components}
\end{figure}

\autoref{fig:r63_components} shows the evolution of the radii of gas and dust in the primary disc over the course of the encounter. We follow \citet{Bate_2018} and \citet{Cuello_2019} in defining the disc's characteristic radius as the radius at which 63.2\% of the disc mass is enclosed. We can again see the significant truncation of the dust disc after the prograde flybys. We also observe a decrease in the radius of the dust disc \emph{prior} to the flyby while the radius of the gas disc stays constant, and a decrease in dust radius in the \texttt{no\_flyby} case. This can again be attributed to radial drift of the dust.

\subsection{Tracking particles in dust and gas maxima} \label{sec:tracking_results}

We utilise the procedure outlined in \autoref{sec:dust_tracking_algorithm} in order to study the behaviour of dust particles in both dust and gas maxima over time. Because of the numerical artefact observed in \autoref{fig:dust_snapshots}, where the dust density becomes unphysically high in the inner regions of the disc, for each simulation we select only particles that are not in the innermost part of the disc (i.e. $r<35$ AU) at the end of the encounter.

\subsubsection{Dust substructures}

\begin{figure}
    \centering
    \includegraphics{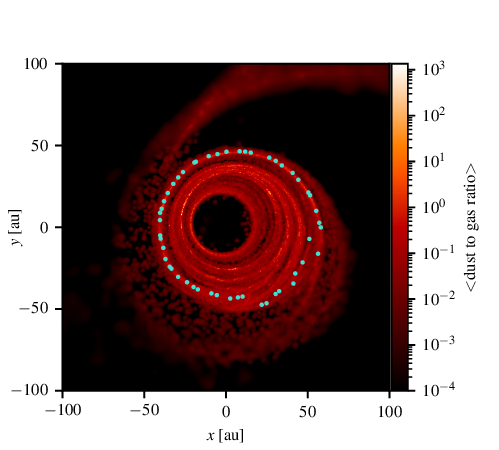}
    \caption{Column-integrated view of the dust to gas ratio at $t=0.6$ (5306 yr) in the \texttt{standard\_run} simulation, with the locations of the 50 tracked dust particles in the dust rings marked in blue.}
    \label{fig:sr_dust_tracked_map}
\end{figure}

\begin{figure}
    \centering
    \begin{subfigure}{\columnwidth}
        \includegraphics{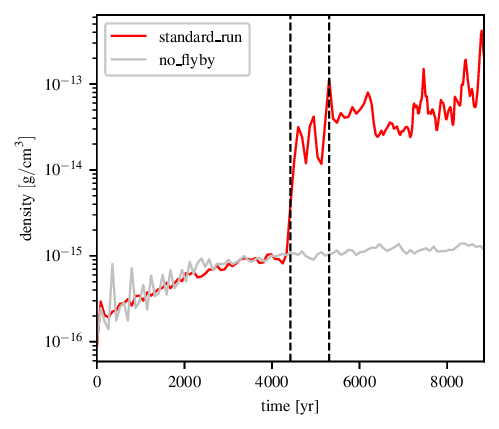}
        \caption{Dust density of tracked particles as a function of time.}
        \label{fig:sr_dust_tracked_rho}
    \end{subfigure}
    \begin{subfigure}{\columnwidth}
        \includegraphics{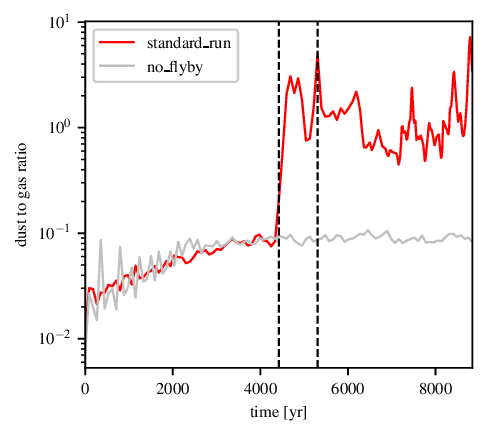}
        \caption{Dust to gas ratio of tracked particles as a function of time.}
        \label{fig:sr_dust_tracked_dtg}
    \end{subfigure}
    \caption{Mean (\subref{fig:sr_dust_tracked_rho}) dust density and (\subref{fig:sr_dust_tracked_dtg}) dust to gas ratio of  50 dust particles tracked from dust substructures in the \texttt{standard\_run} simulation (red line), and a comparison set of 50 dust particles with initial radius $r = 60 \text{ AU}$ in the unperturbed \texttt{no\_flyby} disc (grey line). The dashed lines mark $t=0.5$ and $t=0.6$ respectively for the \texttt{standard\_run} simulation.}
    \label{fig:sr_dust_tracking}
\end{figure}

We first study the rings of high dust to gas ratio induced by the flyby. \autoref{fig:sr_dust_tracked_map} shows the locations of fifty SPH dust particles selected from a dust maximum in the \texttt{standard\_run} simulation. We study the dust maximum at $t=0.6$ because, as can be seen from \autoref{fig:dust_snapshots}, by this time the substructures in the dust disc are clearly defined in the \texttt{standard\_run} simulation and have not yet dissipated. \autoref{fig:sr_dust_tracking} shows the mean dust density and dust to gas ratio of these particles over time. The flyby causes a sudden increase in the particles' dust density and dust to gas ratio of around two orders of magnitude; and, most notably, particles at a similar initial radius in the unperturbed \texttt{no\_flyby} disc exhibit no such increase, suggesting that the flyby itself causes local dust enhancement in the disc. The increases in density and dust to gas ratio persist until the end of the simulation, suggesting that the dust substructures act as bona fide dust traps as opposed to temporary ``traffic jams'' (in which case we would expect to see a temporary peak in the dust to gas ratio around $t=0.6$ followed by a decrease once the tracked particles have moved out of the dust maximum).

\begin{figure}
    \centering
    \begin{subfigure}{\columnwidth}
        \includegraphics{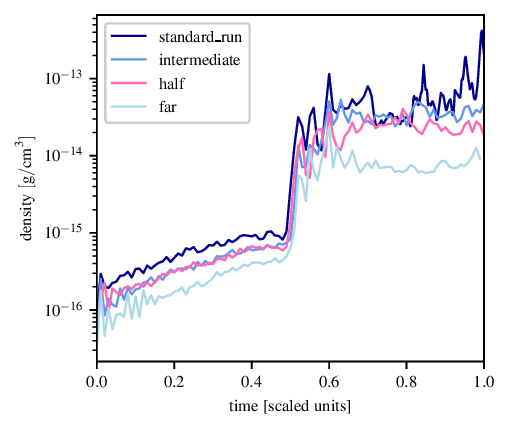}
        \caption{Dust density of tracked particles as a function of time.}
        \label{fig:pg_dust_tracked_rho}
    \end{subfigure}
    \begin{subfigure}{\columnwidth}
        \includegraphics{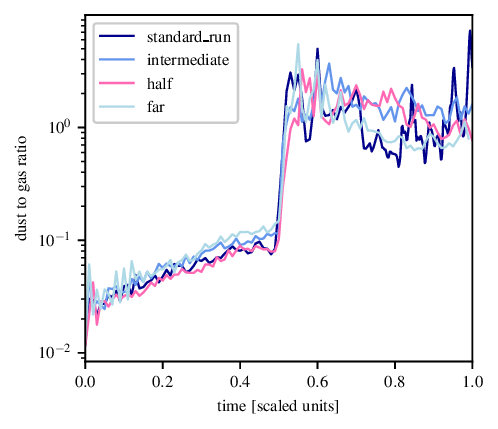}
        \caption{Dust to gas ratio of tracked particles as a function of time.}
        \label{fig:pg_dust_tracked_dtg}
    \end{subfigure}
    \caption{Mean (\subref{fig:pg_dust_tracked_rho}) dust density and (\subref{fig:pg_dust_tracked_dtg}) dust to gas ratio of fifty particles tracked from dust maxima at $t=0.6$ in each of the prograde encounters.}
    \label{fig:pg_dust_tracking}
\end{figure}

In \autoref{fig:pg_dust_tracking} we repeat this procedure for the other three prograde simulations (\texttt{intermediate}, \texttt{half} and \texttt{far}), in each case selecting 50 dust particles from the dust substructures at $t=0.6$ and following their mean dust density and dust to gas ratio over time. We observe extremely similar behaviour in all four cases: at $t=0.5$ there is a sudden increase of about two orders of magnitude in both the dust density and dust to gas ratio of the tracked particles, with the dust to gas ratio reaching unity in the dust maxima. As we observe in \autoref{fig:sr_dust_tracking}, the enhancement in dust density and dust to gas ratio persists for long after periastron, once again suggesting that the flyby-induced substructures in the dust distribution act as dust traps rather than traffic jams. The enhancement in the tracked particles' dust density is weaker in the \texttt{far} simulation compared to the other three encounters, suggesting that more non-penetrating flybys induce less efficient dust traps.

\begin{figure}
    \centering
    \begin{subfigure}{\columnwidth}
        \includegraphics{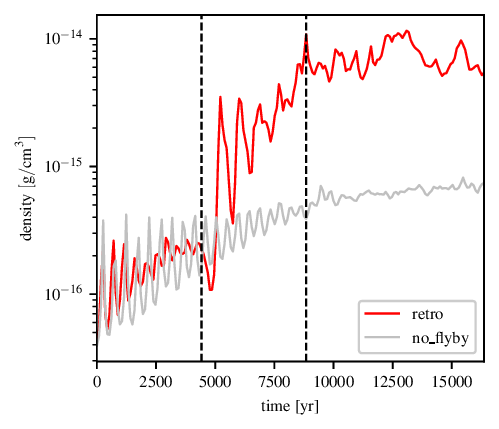}
        \caption{Dust density of tracked particles as a function of time.}
        \label{fig:retro_dust_tracked_rho}
    \end{subfigure}
    \begin{subfigure}{\columnwidth}
        \includegraphics{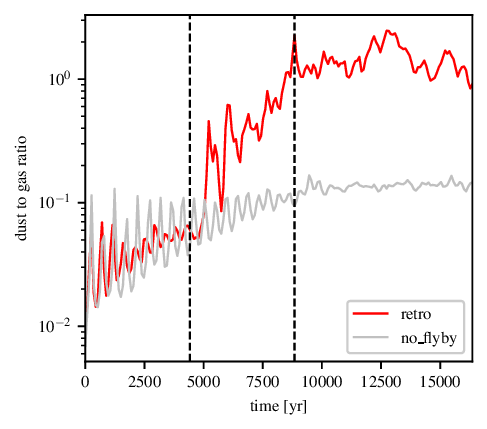}
        \caption{Dust to gas ratio of tracked particles as a function of time.}
        \label{fig:retro_dust_tracked_dtg}
    \end{subfigure}
    \caption{Mean (\subref{fig:retro_dust_tracked_rho}) dust density and (\subref{fig:retro_dust_tracked_dtg}) dust to gas ratio of 50 dust particles tracked from dust substructures in the \texttt{retro} simulation (red line), and a comparison set of 50 dust particles with initial radius $r = 100 \text{ AU}$ in the unperturbed \texttt{no\_flyby} disc (grey line). The dashed lines mark $t=0.5$ and $t=1$ respectively for the \texttt{retro} simulation.}
    \label{fig:retro_dust_tracking}
\end{figure}

We then perform a similar exercise for the retrograde simulations. As \autoref{fig:dust_snapshots} shows, dust substructures persist for longer in the retrograde encounters than they do for prograde encounters. We therefore choose to track particles from dust rings $t=1$ rather than at $t=0.6$, reasoning that if the substructures have persisted for longer they may be more likely to exhibit dust trapping. \autoref{fig:retro_dust_tracking} shows the dust density and dust to gas ratio as a function of time of 50 SPH particles tracked from a dust maximum at $t=1$ in the \texttt{retro} simulation. Again, we observe a sudden dust enhancement at $t=0.5$ that persists for the rest of the simulation, and an increase in dust to gas ratio of an order of magnitude more than is observed in a comparison set of dust particles from the \texttt{no\_flyby} case. (Note that the in-phase oscillations in dust density and dust to gas ratio of the \texttt{no\_flyby} particles is a numerical artefact caused by the dust particles having no initial vertical velocity in the simulation set-up; the dust disc has relaxed by the time of the flyby and so this artefact has no bearing on the \texttt{no\_flyby} particles' dust density and dust to gas ratio at later times.) The enhancement in dust density and dust to gas ratio of the tracked particles is weaker in the \texttt{retro} simulation than in the prograde encounters, although still significant compared to the \texttt{no\_flyby} case; this less efficient trapping of dust particles can be attributed to the reduced interaction time between the disc particles and the perturber during a retrograde encounter due to their higher relative velocity. In Appendix \ref{sec:no_trapping} we show that increasing the periastron distance of the retrograde encounter, as in the \texttt{int\_retro} simulation, results in the dust substructures failing to trap dust particles at all.

\subsubsection{Gas substructures}

Following the suggestions by \citet{Cuello_2019} and \citet{Smallwood_2023} that the spiral overdensities in \emph{gas} induced by flybys could also act as dust traps, we now perform the same procedure on the gas maxima in each disc. Although from \autoref{fig:gas_snapshots} we observe spiral arms forming a bridge of material between the two stars at periastron for each of the prograde simulations, we are not able to select a statistically significant number of SPH dust particles from the bridge that remain in the primary disc at the end of the encounter in most cases, and so we do not perform any tracking from $t=0.5$. 

\begin{figure}
    \centering
    \includegraphics{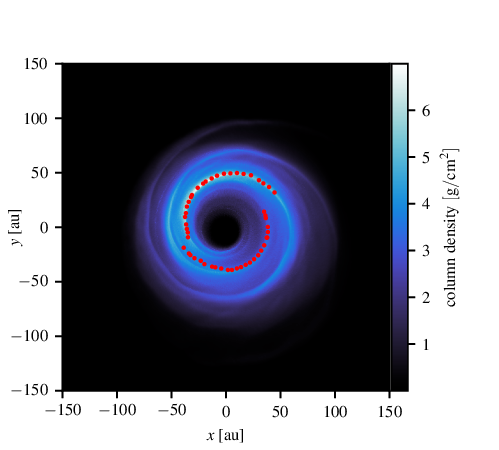}
    \caption{Column density at $t=0.6$ (9747 yr) in the \texttt{intermediate} simulation, with the locations of the 50 tracked dust particles in the gas spirals marked in red.}
    \label{fig:intermediate_gas_tracked_map}
\end{figure}

\begin{figure}
    \centering
    \begin{subfigure}{\columnwidth}
        \includegraphics{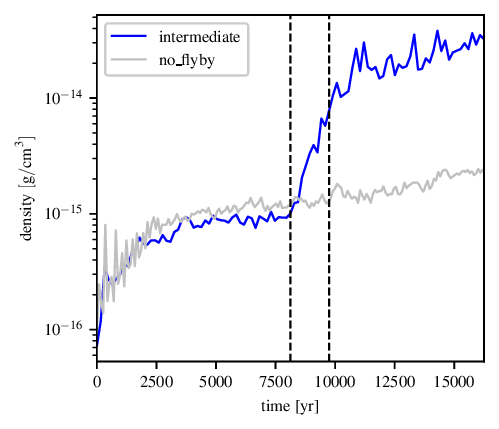}
        \caption{Dust density of tracked particles as a function of time.}
        \label{fig:intermediate_gas_tracked_rho}
    \end{subfigure}
    \begin{subfigure}{\columnwidth}
        \includegraphics{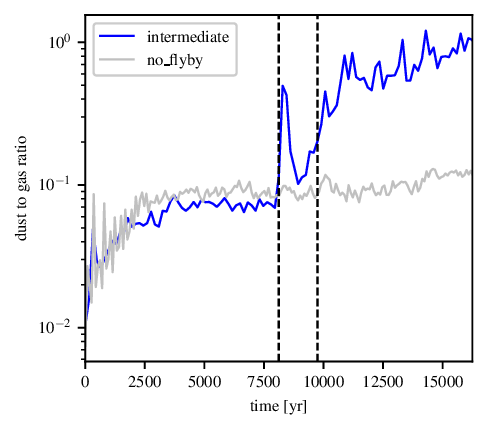}
        \caption{Dust to gas ratio of tracked particles as a function of time.}
        \label{fig:intermediate_gas_tracked_dtg}
    \end{subfigure}
    \caption{Mean (\subref{fig:intermediate_gas_tracked_rho}) dust density and (\subref{fig:intermediate_gas_tracked_dtg}) dust to gas ratio of 50 dust particles tracked from gas spirals in the \texttt{intermediate} simulation (blue line), and a comparison set of 50 dust particles with initial radius $r = 60 \text{ AU}$ in the unperturbed \texttt{no\_flyby} disc (grey line). The dashed lines mark $t=0.5$ and $t=0.6$ respectively for the \texttt{intermediate} simulation.}
    \label{fig:intermediate_gas_tracking}
\end{figure}

We instead once again select particles from the gas spirals at $t=0.6$, as shown in \autoref{fig:intermediate_gas_tracked_map}. \autoref{fig:intermediate_gas_tracking} shows the evolution of these particles' dust density and dust to gas ratio over time. Once again, we see an increase in dust density and dust to gas ratio of around 1 order of magnitude above those of a comparison set of particles in the \texttt{no\_flyby} disc. In \autoref{fig:intermediate_gas_tracked_dtg} we observe a sharp peak in dust to gas ratio of about 0.5 at $t=0.51$, which then decreases before slowly increasing again over time. There is no corresponding peak in dust density at $t=0.51$ in \autoref{fig:intermediate_gas_tracked_rho}, so this peak cannot be interpreted as a temporary ``traffic jam'' in which the dust particles are briefly in a region of high dust density. Instead, a few of the 50 tracked dust particles are briefly in a region of very low \emph{gas} density at $t=0.51$, which results in a temporary peak in the mean dust to gas ratio.

\begin{figure}
    \centering
    \begin{subfigure}{\columnwidth}
        \includegraphics{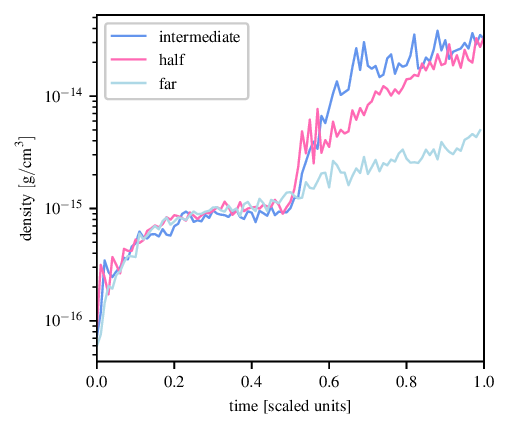}
        \caption{Dust density of tracked particles as a function of time.}
        \label{fig:pg_gas_tracked_rho}
    \end{subfigure}
    \begin{subfigure}{\columnwidth}
        \includegraphics{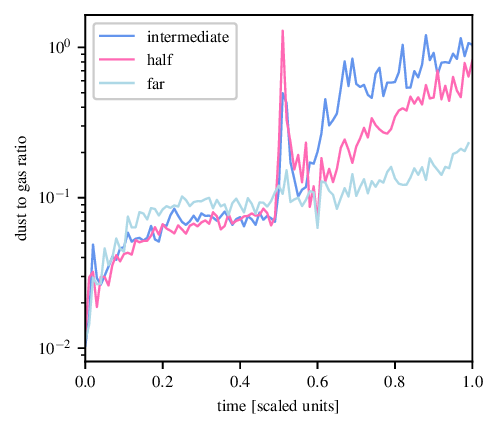}
        \caption{Dust to gas ratio of tracked particles as a function of time.}
        \label{fig:pg_gas_tracked_dtg}
    \end{subfigure}
    \caption{Mean (\subref{fig:pg_gas_tracked_rho}) dust density and (\subref{fig:pg_gas_tracked_dtg}) dust to gas ratio of fifty particles tracked from gas maxima at $t=0.6$ in the prograde \texttt{intermediate}, \texttt{far} and \texttt{half} simulations.}
    \label{fig:pg_gas_tracking}
\end{figure}

In \autoref{fig:pg_gas_tracking} we repeat this procedure for the other prograde simulations, with the exception of the \texttt{standard\_run} simulation, because as can be seen in \autoref{fig:gas_snapshots} the spiral arms in this encounter dissipate too quickly to be able to perform meaningful particle tracking. In each case, we select 50 dust particles from the gas spiral arms at $t=0.6$ and follow their mean dust density and dust to gas ratio over time. We observe fairly similar behaviour between the \texttt{intermediate} and \texttt{half} simulations, with an overall increase of about 2 orders of magnitude in both dust density and dust to gas ratio of the tracked particles over the course of the flyby encounter. The \texttt{half} encounter has an even more pronounced peak in dust to gas ratio at $t=0.51$, which can again be explained by a few of the tracked particles being briefly in a region of very low gas density at that time.

However, the tracked particles in the \texttt{far} simulation behave differently, with no peak at $t=0.5$ and an increase of only an order of magnitude in both dust density and dust to gas ratio over the course of the flyby. In Appendix \ref{sec:no_trapping} we study the behaviour of these particles in more detail and compare it to that of particles in the unperturbed disc. We find that there is \textbf{no} dust trapping in the gas spiral arms of the \texttt{far} simulation, and their smaller increase in dust density over the course of the flyby can be attributed to radial drift.

\begin{figure}
    \centering
    \begin{subfigure}{\columnwidth}
        \includegraphics{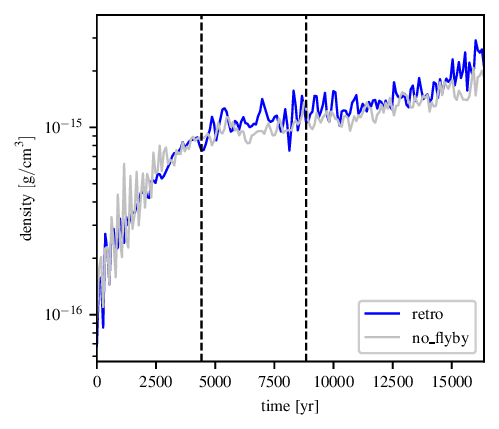}
        \caption{Dust density of tracked particles as a function of time.}
        \label{fig:retro_gas_tracked_rho}
    \end{subfigure}
    \begin{subfigure}{\columnwidth}
        \includegraphics{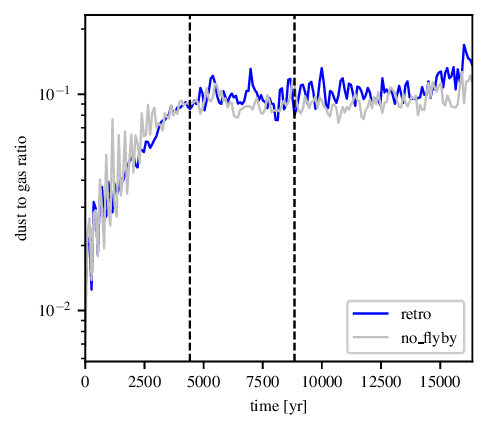}
        \caption{Dust to gas ratio of tracked particles as a function of time.}
        \label{fig:retro_gas_tracked_dtg}
    \end{subfigure}
    \caption{Mean (\subref{fig:retro_gas_tracked_rho}) dust density and (\subref{fig:retro_gas_tracked_dtg}) dust to gas ratio of 50 dust particles tracked from gas spirals in the \texttt{retro} simulation (blue line), and a comparison set of 50 dust particles with initial radius $r = 65 \text{ AU}$ in the unperturbed \texttt{no\_flyby} disc (grey line). The dashed lines mark $t=0.5$ and $t=1$ respectively for the \texttt{retro} simulation.}
    \label{fig:retro_gas_tracking}
\end{figure}

Once again, we can perform a similar exercise for the retrograde simulations. \autoref{fig:gas_snapshots} shows that the flyby-induced spiral arms last for considerably longer in the retrograde encounters than in the prograde encounters, so we choose to track particles in the gas spiral arms at $t=1$ rather than $t=0.6$, reasoning once again that if the spiral structures have persisted for longer they may be more likely to exhibit dust trapping. In \autoref{fig:retro_gas_tracking} we show the behaviour over time of fifty dust particles selected from the gas spirals at $t=1$  in the \texttt{retro} simulation. Although the gas spiral arms persist for many thousands of years after the retrograde encounter, we observe no difference between the tracked particles and a comparison set of particles from the unperturbed disc, suggesting that there is \textbf{no} dust trapping in the gas spiral arms in the retrograde encounters. This can again be attributed to the weaker interaction between a retrograde perturber and the disc.

\subsubsection{Effect of varying flyby parameters}

\begin{table}
\centering
\caption{A summary of particle tracking results by flyby inclination. "Yes" indicates that dust trapping was found in every case, "sometimes" that dust trapping was found in some cases, and "no" indicates that no dust trapping was found.}
\label{tab:tracking_results}
\begin{tabular}{ccc}
    \hline
    inclination & dust substructures & gas substructures \\ \hhline{===}
    prograde & yes & sometimes \\ \hline
    retrograde & sometimes & no \\ \hline
\end{tabular}
\end{table}

\autoref{tab:tracking_results} summarises the particle tracking results presented in \autoref{sec:tracking_results}. We have found evidence of dust trapping in the flyby-induced dust substructures in all the prograde encounters as well as the \texttt{retro} encounter, but \emph{not} the \texttt{int\_retro} encounter. On the other hand, flyby-induced gas substructures are less efficient at trapping dust, as we only observe dust trapping for sufficiently close-in prograde encounters and in neither of the retrograde encounters.

The most important factor in whether or not a flyby can induce dust-trapping substructures in the perturbed disc is therefore its orientation; retrograde flybys, due to their weaker interactions with the disc, induce substructures that do not trap dust as strongly. 

Increasing the periastron distance of the flyby also leads to a weaker interaction with the disc and hence less efficient dust trapping; the \texttt{far} simulation shows weaker dust trapping than the \texttt{intermediate} simulation in dust substructures (see \autoref{fig:pg_dust_tracking}) and no dust trapping at all in the flyby-induced gas spirals (see \autoref{fig:far_gas_tracking}). Similarly, when we compare the \texttt{retro} and \texttt{int\_retro} simulations, we find dust trapping in the dust substructures for \texttt{retro} (\autoref{fig:retro_dust_tracking}) but not for the same substructures in the \texttt{int\_retro} simulation (see \autoref{fig:ir_dust_tracking}). 

We also varied the mass of the perturber from $M_{\sun}$ to $0.5M_{\sun}$ in the \texttt{half} simulation. This has much the same effect as increasing the periastron distance of the flyby, as we see from Figures \ref{fig:pg_dust_tracking} and \ref{fig:pg_gas_tracking} that the \texttt{half} and \texttt{intermediate} simulations behave very similarly. We can therefore conclude that grazing prograde flybys with a range of perturber masses are likely to induce dust trapping in the perturbed disc; however, only sufficiently close-in retrograde flybys are able to induce dust trapping.

\section{Discussion} \label{sec:discussion}

\subsection{Relevance for the streaming instability}

Although we have shown that flyby-induced substructures cause an increase in tracked particles' dust to gas ratio that persists for long periods of time, it remains to be determined whether the disc's response to the flyby can induce planetesimal formation. We therefore now investigate whether flybys can trigger the streaming instability.

\citetalias{Li_Youdin_2021} ran a suite of SI simulations for a range of \emph{surface} (i.e. vertically integrated) dust to gas ratios $Z$ and particle Stokes numbers St in order to find the threshold value of $Z$ required to to produce strong particle clumping and trigger the SI as a function of St. They produced an analytical fit to their clumping boundary (for $0.015 < \text{St} < 1$)

\begin{equation}
    \log\left({\frac{Z_\mathrm{crit}}{\Pi}}\right) = 0.13(\log{\mathrm{St}})^2 + 0.1(\log{\mathrm{St}}) -1.07,
    \label{eqn:z_crit}
\end{equation}

\noindent where $\Pi$ is a dimensionless parameter related to the radial pressure gradient in the disc, which takes the value $\Pi\approx0.05$ in our simulations. The SI is triggered in a disc if the local value of $Z$ exceeds $Z_\text{crit}$. We therefore convert the dust to gas ratio, $\rho_d/\rho_g$, of each of our tracked particles (as shown in \autoref{sec:tracking_results}) to a surface dust to gas ratio $Z$ using

\begin{equation}
    Z = \frac{\rho_d}{\rho_g}\frac{H_d}{H_g}
    \label{eqn:surface_dtg}
\end{equation}

\noindent where $H_d$ and $H_g$ are the scale heights of dust and gas respectively in the disc. We then use \autoref{eqn:stokes} to compute the Stokes number for each tracked particle as a function of time, from which we can use \autoref{eqn:z_crit} to find the critical value $Z_\text{crit}$ that must be exceeded in order to trigger the SI. 
 
\begin{figure}
    \centering
    \includegraphics{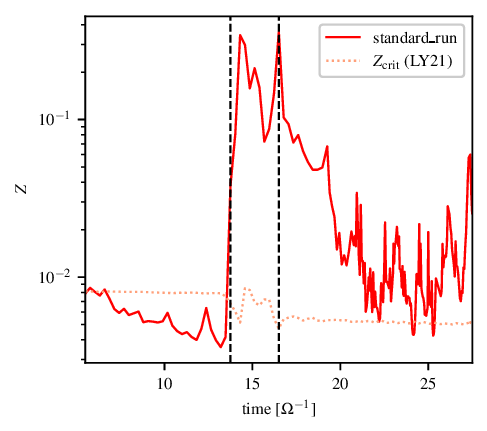}
    \caption{Solid red line: mean surface dust to gas ratio $Z$, as calculated using \autoref{eqn:surface_dtg}, of the 50 tracked particles from dust substructures in the \texttt{standard\_run} simulation shown in \autoref{fig:sr_dust_tracked_map}. Dotted light red line: mean value of $Z_\text{crit}$ calculated for each particle as a function of time using \autoref{eqn:z_crit}. We scale the time axis by the dynamical time $\Omega^{-1}$ at the position of the tracked particles. The dashed lines mark $t=0.5$ and $t=0.6$ respectively for the \texttt{standard\_run} simulation. Note that the plot starts a little after $t=0$ in order to give the dust particles time to settle to the midplane.}
    \label{fig:surface_dtg}
\end{figure}

\autoref{fig:surface_dtg} shows the mean surface dust to gas ratio of the tracked particles from the \texttt{standard\_run} dust substructures (shown in \autoref{fig:sr_dust_tracked_map}) and the threshold value of $Z$ needed to trigger the SI as a function of time. We see that after the flyby, the threshold is exceeded until the end of the simulation, suggesting that the streaming instability \emph{can} be triggered in flyby-induced dust substructures. 

The tracked particles are in a dust substructure with radius $r\approx50$ AU at $t=0.6$, and so we choose to scale the time axis in \autoref{fig:surface_dtg} by the dynamical time at this radius, $\Omega^{-1}\approx 320$ yr. We can therefore see that the tracked particles' surface dust to gas ratio exceeds the threshold for SI for many tens of dynamical times after the flyby. As the SI  is induced on timescales of the order of one dynamical time (see \citealt{pp7} and references therein), this further suggests that the particles remain in regions of high dust to gas ratio for long enough time periods to trigger planetesimal formation via the SI.

In Appendix \ref{sec:si_more} we perform a similar exercise for the other five flyby encounters and find that in all five cases where dust trapping is observed in the dust substructures, the threshold value of $Z$ required for SI is exceeded by the tracked particles. However, in the two simulations where dust trapping was observed in the \emph{gas} spiral arms, \texttt{intermediate} and \texttt{half}, we find that the threshold value of $Z$ required to trigger SI is \textbf{not} exceeded by particles tracked from the gas spirals.

We also note that in all four of the prograde simulations, the value of $Z$ for the tracked particles rises to a peak at around $t=0.6$ before declining again. Notably, the value of the dust to gas ratio $\rho_d/\rho_g$ itself does not decline after the flyby, as shown in Figures \ref{fig:sr_dust_tracked_dtg} and \ref{fig:pg_dust_tracked_dtg}; from \autoref{eqn:surface_dtg} we therefore conclude that the dimensionless factor $H_d/H_g$ decreases over time for the prograde simulations, but does not do so in the case of the \texttt{retro} simulation. We attribute this difference to a prograde flyby causing greater perturbation to the dust disc and accelerating dust settling to the midplane, which hence results in a smaller value of the dust scale height $H_d$ in the prograde case and a smaller value of the surface dust to gas ratio $Z$.

We have therefore shown that solar-mass prograde perturbers can potentially trigger SI for a range of periastron distances at least as large as $r_\text{peri}=350$ AU in the case of the large discs (outer radius 150 AU) simulated here. As a stellar flyby is most commonly defined to include any unbound encounter within 1000 AU \citep{Cuello_Ménard_Price_2023}, it is worth considering whether or not these relatively close-in flybys are sufficiently common to be a statistically significant mechanism for planetesimal formation.

\citet{Clarke_Pringle_1991} estimated the rate $\Gamma$ of equal-mass flyby encounters as a function of stellar environment. They found

\begin{equation}
    \Gamma = \frac{4\sqrt{\pi}n_0GMr_\mathrm{peri}}{V_*}\left(1+\frac{V_*^2r_\mathrm{peri}}{GM}\right),
    \label{eqn:encounter_rates}
\end{equation}

\noindent where $n_0$ is the number density of stars in the environment, $M$ is the mass of either star in an equal-mass encounter and $V_*$ is the one-dimensional stellar velocity dispersion. For a dense young cluster such as the inner region of the Orion Nebula Cluster, $n_0=2 \times 10^4 \mathrm{pc}^{-3}$ \citep{Marks_Kroupa_2012} and $V_*=2.6 \mathrm{km/s}$ \citep{Stutz_2017}, \autoref{eqn:encounter_rates} yields an encounter rate with $r_\mathrm{peri}=350$ AU of $\Gamma=1.5 \mathrm{Myr}^{-1}$, corresponding to a significant number of young discs in dense stellar clusters that might undergo a close-in flyby. However, it is important to note that in these dense regions, the far-ultraviolet flux is large enough that external photoevaporation would destroy a disc of extent 150 AU on a shorter timescale than a flyby encounter \citep{Winter_Clarke_Rosotti_Ih_Facchini_Haworth_2018}. A lower number density $n_0=1000 \mathrm{pc}^{-3}$ and a ``standard'' velocity dispersion of $V_*=4\mathrm{km/s}$ (\citeauthor{Winter_Clarke_Rosotti_Ih_Facchini_Haworth_2018}) yields a much smaller encounter rate $\Gamma=0.1 \mathrm{Myr}^{-1}$. This suggests that we expect our flyby-triggered dust trapping mechanism to play a significant role in planetesimal formation mainly at stellar number densities higher than $1000 \mathrm{pc}^{-3}$. 

It is worth noting that several recently observed flyby candidates have been found in regions of very low stellar density. \citet{Menard_2020} studied the UX Tauri system, which is found in the low-density Taurus constellation, with $n=1-10\mathrm{pc}^{-3}$. They suggested that UX Tau A is undergoing a flyby by UX Tau C with periastron distance $r_\mathrm{peri}=200$ AU, well within the distance required to see dust trapping in the flyby-induced spiral arms. \citeauthor{Menard_2020} suggest this could be a result of the ``patchiness'' of Taurus and the flyby rate is higher in denser groups. RW Aurigae, which \citet{Dai_2015} have modelled as a flyby with \emph{current}
separation 200 AU (suggesting the periastron distance of the flyby is even closer) is similarly located in the Taurus-Auriga complex. It is therefore clear that sufficiently close-in flybys can occur and have been observed even in regions of low stellar density, at a rate that we would not predict using \autoref{eqn:encounter_rates}. These candidates may in fact be very loosely bound binaries rather than unbound flyby encounters, or else reflect the inhomogeneous stellar distributions in star-forming regions. These order-of-magnitude calculations do, however, suggest that flybys with a relatively small periastron distance such that they can induce dust trapping via the SI are not uncommon in star-forming regions.

As a caveat, this is a proof-of-concept treatment; 
to fully understand whether these conditions can lead to the SI, we would need to input the local dust to gas ratios and Stokes numbers at the location of the flyby-induced substructures into 
a 2D shearing box simulation, but we have shown here that flybys \emph{do} locally enhance the surface dust to gas ratio to values large enough to create favourable conditions for the SI.

\subsection{Effect of dust grain size} \label{sec:grain_size}

For computational reasons, the simulations presented in this work all have a dust grain size $s = 3$ cm. However, it is useful to examine dust particles with different Stokes numbers in order to determine whether the flyby-induced enhancement in dust to gas ratio is reproduced in different dust species. \autoref{eqn:z_crit} suggests that the value of $Z_\mathrm{crit}$ has a minimum at $\mathrm{St} \sim 0.3$, and so dust grains with lower Stokes numbers than those considered here are particularly pertinent to this question.

\begin{table}
\centering
\caption{Simulations used for dust grain size comparison.}
\label{tab:1cm_simulations}
\begin{tabular}{cccc}
    \hline
    Simulation ID & $M_2$ [$M_{\sun}$] & $r_{\text{peri}}$ [au] & flyby inclination \\ \hhline{====}
    standard\_run & 1.0 & 175.0 & prograde \\ \hline
    pg\_1cm & 1.0 & 200.0 & prograde \\ \hline
    retro & 1.0 & 175.0 & retrograde \\ \hline
    rg\_1cm & 1.0 & 200.0 & retrograde \\ \hline
\end{tabular}
\end{table}

We therefore compare our results to the two-fluid coplanar stellar flyby simulations presented by \citet{Cuello_2019}. \texttt{pg\_1cm} (their \texttt{beta0}) is a prograde flyby with 1cm dust grains and a periastron distance $r_\mathrm{peri}=200$ AU. \texttt{rg\_1cm} (their \texttt{beta180}) is a retrograde flyby with 1cm dust grains and a periastron distance $r_\mathrm{peri}=200$ AU. All other disc and perturber parameters are identical to those of the \texttt{standard\_run} and \texttt{retro} simulations respectively, as summarised in \autoref{tab:1cm_simulations}.

\begin{figure}
    \centering
    \begin{subfigure}{\columnwidth}
        \includegraphics{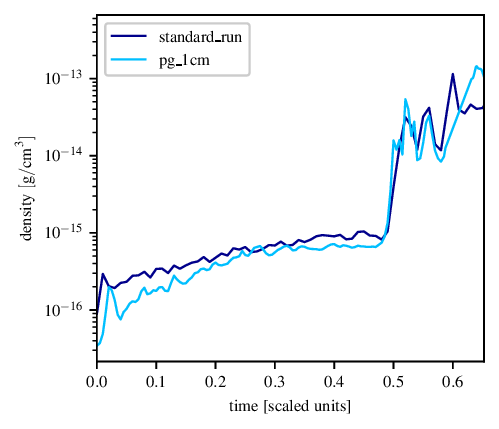}
        \caption{Dust density of tracked particles as a function of time.}
        \label{fig:pg_rho_sizecomp}
    \end{subfigure}
    \begin{subfigure}{\columnwidth}
        \includegraphics{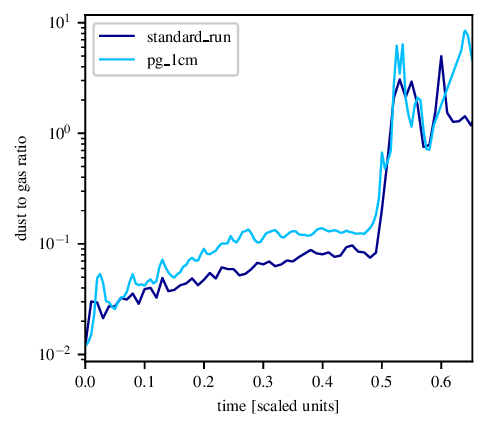}
        \caption{Dust to gas ratio of tracked particles as a function of time.}
        \label{fig:pg_dtg_sizecomp}
    \end{subfigure}
    \caption{Mean (\subref{fig:pg_rho_sizecomp}) dust density and (\subref{fig:pg_dtg_sizecomp}) dust to gas ratio of fifty particles tracked from dust maxima at $t=0.6$ in the \texttt{standard\_run} and \texttt{pg\_1cm} simulations.}
    \label{fig:pg_sizecomp}
\end{figure}

\begin{figure}
    \centering
    \begin{subfigure}{\columnwidth}
        \includegraphics{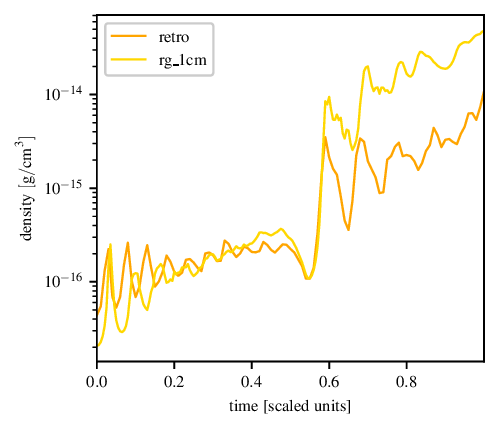}
        \caption{Dust density of tracked particles as a function of time.}
        \label{fig:rg_rho_sizecomp}
    \end{subfigure}
    \begin{subfigure}{\columnwidth}
        \includegraphics{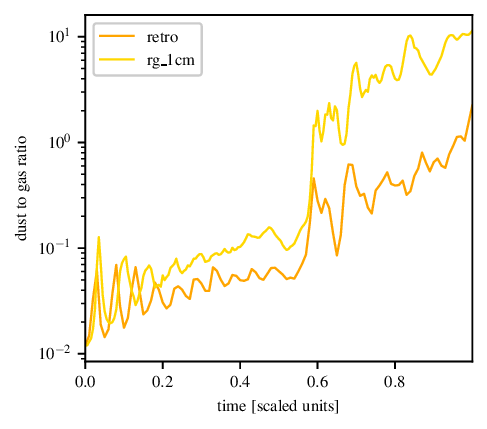}
        \caption{Dust to gas ratio of tracked particles as a function of time.}
        \label{fig:rg_dtg_sizecomp}
    \end{subfigure}
    \caption{Mean (\subref{fig:rg_rho_sizecomp}) dust density and (\subref{fig:rg_dtg_sizecomp}) dust to gas ratio of fifty particles tracked from dust maxima at $t=1$ in the \texttt{retro} and \texttt{rg\_1cm} simulations.}
    \label{fig:rg_sizecomp}
\end{figure}

\autoref{fig:pg_sizecomp} shows the behaviour over time of 50 dust particles selected from the dust substructures at $t=0.6$ in the \texttt{standard\_run} (as shown in Figures \ref{fig:sr_dust_tracked_map} and \ref{fig:sr_dust_tracking}) and \texttt{pg\_1cm} simulations. We see that the smaller dust particles in the \texttt{pg\_1cm} simulation actually have enhancements in their dust to gas ratio even higher than those of the \texttt{standard\_run} particles, reaching a mean dust to gas ratio of nearly 10 soon after the flyby. Similarly, \autoref{fig:rg_sizecomp} shows the behaviour over time of 50 dust particles selected from the dust substructures at $t=1$ in both the \texttt{retro} and \texttt{rg\_1cm} simulations. Once again, the flyby-induced enhancement in dust density and dust to gas ratio is even higher in the 1cm dust grain simulation. Notably, although we have earlier seen that increasing the periastron distance of the flyby weakens the strength of dust trapping, the slight increase in periastron distance of the 1cm dust grain simulations is entirely cancelled out by the effect of decreasing dust grain size.

It is also important to consider the effect of a polydisperse dust grain size distribution on the efficacy of flyby-induced dust trapping. \citet{Krapp_2019} studied the effect on SI growth rates when multiple dust species with different Stokes numbers were present, and found that the SI grows much more slowly in cases with polydisperse dust distributions. Notably, when the maximum Stokes number of the distribution was $\mathrm{St}=1$, they found the SI developed on timescales of at least $10^3\Omega^{-1}$. As Figures \ref{fig:surface_dtg} and \ref{fig:surface_dtg_grid} show, the tracked particles' surface dust to gas ratio does not exceed the threshold required to trigger SI for these timescales. However, \citeauthor{Krapp_2019} note that if dust particles are concentrated in substructures and segregated by particle size, favourable conditions for SI may still develop. As this work examines the potential for SI \emph{within} flyby-induced substructures, it is still possible even in a polydisperse dust grain size distribution that the SI can be triggered in a disc perturbed by a stellar flyby.

\subsection{Observational implications}

\citet{Cuello_2020} identified several observational signatures which could be used to identify stellar flyby candidates, namely flyby-induced spirals, stellar bridges and disc truncation (as well as warps and disc misalignment for inclined flybys). In order to determine whether or not SI is being triggered in the observed substructures in a flyby candidate disc, as \autoref{fig:surface_dtg} suggests, the technique presented by \citet{Zagaria_Clarke_Booth_Facchini_Rosotti_2023} could be used. This technique uses multi-frequency dust continuum observations to determine the dust surface density profile $\Sigma_\mathrm{dust}$, and CO emission line data to derive the gas-to-dust coupling. These two parameters can then be related to the values of $Z$ and St in the disc substructure. Finally, \autoref{eqn:z_crit} determines whether or not the substructure in question is prone to particle clumping under SI.

As \citeauthor{Zagaria_Clarke_Booth_Facchini_Rosotti_2023} demonstrate with the source HD 163296, coupled gas and dust observations are required to observationally determine whether or not SI is taking place in flyby-induced substructures. Further work will investigate the kinematic structures of flyby-perturbed discs and whether they can be identified using CO emission line data.

\subsection{How do the dust substructures form?}

\begin{figure*}
    \centering
    \includegraphics{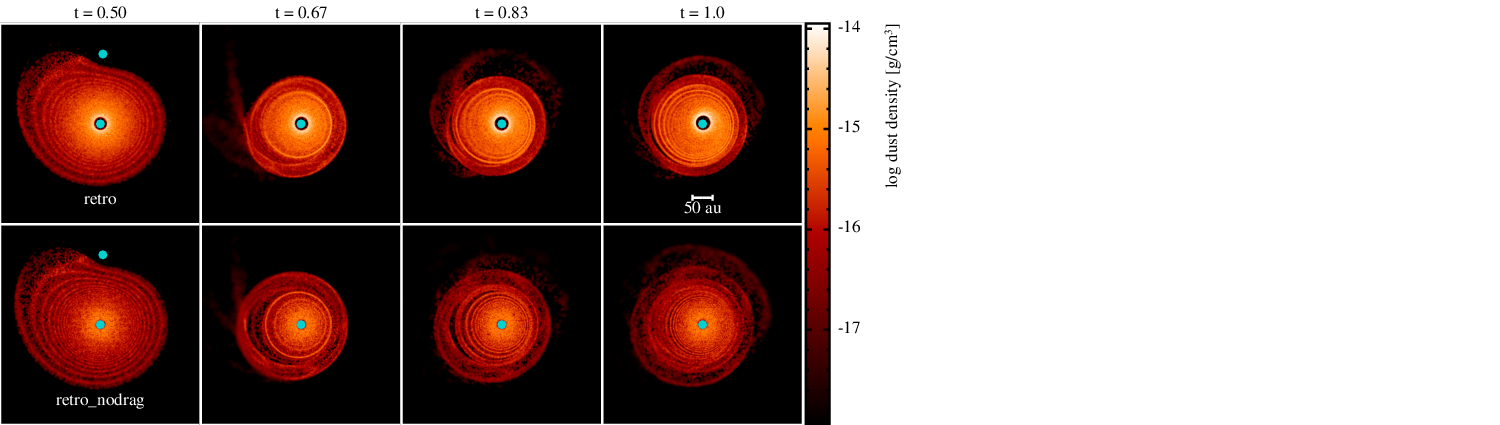}
    \caption{Face-on view of the midplane dust density (i.e. at $z=0$, not column-integrated) of the primary disc for the \texttt{retro} and \texttt{retro\_nodrag} simulations. Columns (left to right) are at times $t=0.5$ (periastron), $t=0.67$, $t=0.83$ and $t=1$. Sink particles (in blue) are large for visualisation purposes only.}
    \label{fig:nodrag_snapshots}
\end{figure*}

From \autoref{fig:dust_snapshots} we see that the flybys induce ring-like substructures in the dust disc that persist for many thousands of years after periastron, and in many cases do not spatially coincide with the gas spirals seen in \autoref{fig:gas_snapshots}. Our results here agree with \citet{Cuello_2019}, who find that cm-sized dust grains do not follow the gas density distribution .

We here focus our attention on the dust substructures in the \texttt{retro} simulation, as these are more prominent and long-lasting than the substructures in any of the prograde encounters. In order to determine whether the substructures in the dust disc are caused by gas-dust interaction or are a purely kinematic effect of the flyby, we ran an additional simulation, \texttt{retro\_nodrag}, with the same initial conditions as the \texttt{retro} simulation but without drag, so that the dust particles behave as N-body test particles.

\autoref{fig:nodrag_snapshots} shows the evolution of the \emph{midplane} dust density in the two cases. We observe the ring-like substructures form in the \texttt{retro\_nodrag} simulation at about $t=0.6$ as they do in the \texttt{retro} simulation, suggesting that the substructures are indeed a kinematic effect of the flyby and do not require gas-dust coupling. However, the midplane dust density of the substructures is much higher in the case with drag than without drag; furthermore, the substructures dissipate much more quickly in the \texttt{retro\_nodrag} case. This implies that although drag forces are not required to form the substructures, they play an important role in maintaining long-lived regions of enhanced dust density.

\begin{figure}
    \centering
    \begin{subfigure}{\columnwidth}
        \includegraphics{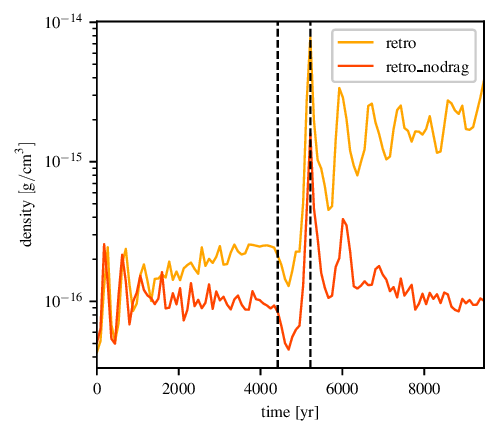}
        \caption{Dust density of tracked particles as a function of time.}
        \label{fig:nodrag_tracked_rho}
    \end{subfigure}
    \begin{subfigure}{\columnwidth}
        \includegraphics{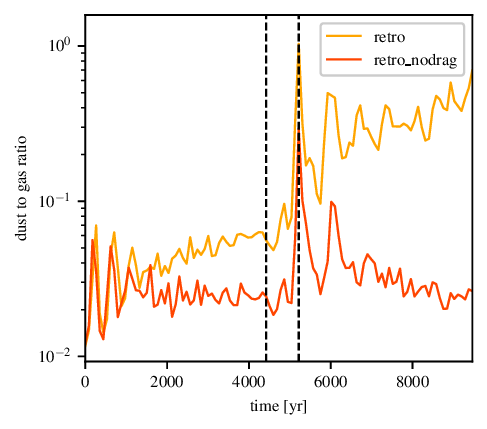}
        \caption{Dust to gas ratio of tracked particles as a function of time.}
        \label{fig:nodrag_tracked_dtg}
    \end{subfigure}
    \caption{Mean (\subref{fig:nodrag_tracked_rho}) dust density and (\subref{fig:nodrag_tracked_dtg}) dust to gas ratio of fifty particles tracked from the dust ring at $t=0.59$ in the \texttt{retro} and \texttt{retro\_nodrag} simulations. The dashed lines mark $t=0.5$ and $t=0.59$ respectively for both simulations.}
    \label{fig:nodrag_tracking}
\end{figure}

We can also study the ability of the dust substructures to trap dust. In \autoref{fig:nodrag_tracking} we track 50 SPH particles from the dust ring formed at $t=0.59$ in both the \texttt{retro} and \texttt{retro\_nodrag} simulations and show their mean dust density and dust to gas ratio as a function of time. The density and dust to gas ratio of the tracked particles remain an order of magnitude higher after the flyby in the \texttt{retro} simulation, but in the \texttt{retro\_nodrag} simulation no significant enhancement or dust trapping is seen. This suggests that, as drag is required to maintain dust overdensities for long periods of time, no dust trapping can take place without gas-dust coupling. As the no-drag case can be considered the $\mathrm{St} \rightarrow \infty$ limit, we can compare these results to those shown in Figures \ref{fig:pg_sizecomp} and \ref{fig:rg_sizecomp} to further confirm that the dust trapping mechanism is most efficient for particles with smaller Stokes numbers.

\section{Conclusions} \label{sec:conclusions}

We have studied the dynamical response of a protoplanetary disc to coplanar stellar flybys using 3D SPH simulations. Our results agree with previous findings that flybys cause tidal stripping, truncation and the formation of spiral arms within the disc.

In this work we study the behaviour of the flyby-induced substructures in detail in order to determine whether or not they behave as dust traps. For each simulation, we track 50 SPH dust particles from within the dust and gas density maxima and compare their behaviour to particles at comparable radii in an unperturbed case. 

We find that particles tracked from \emph{dust} substructures show signs of dust trapping (i.e. a significant enhancement in dust density and dust to gas ratio compared to particles in the unperturbed disc) in both prograde and retrograde flyby encounters. The increase in local dust to gas ratio of the tracked particles persists for many thousands of years after the flyby, suggesting that the substructures act as dust traps and not merely traffic jams.

However, we find that the gas spiral arms only act as dust traps in the case of close-in prograde flybys, which suggests that a sufficiently strong interaction with the perturber is required for the pressure maxima in the gas disc to act as dust traps.

We also compare the local dust to gas ratios induced by the flyby to the critical surface dust to gas ratios required to trigger planetesimal formation via the streaming instability. We find that the surface dust to gas ratios of the tracked particles in dust substructures exceed the threshold required to trigger SI after the flyby and remain high for many dynamical timescales after the flyby. This implies that stellar flybys are a potential mechanism for inducing the SI and planetesimal formation.

\section*{Acknowledgements}

VRP thanks the Science and Technology Facilities Council (STFC)
for a Ph.D. studentship. CL and CJC have been supported by the UK Science and Technology research Council (STFC) via the consolidated grant ST/W000997/1.

The authors thank Nicolás Cuello for the use of \textsc{Phantom} flyby simulation data in \autoref{sec:grain_size}, Stéphane Michoulier for implementing the implicit drag scheme in \textsc{Phantom}, Daniel Price and Bec Nealon for helpful discussions and the anonymous reviewer for constructive comments.

The simulations presented in this work were performed using the DiRAC Data Intensive service at Leicester (DiAL3), operated by the University of Leicester IT Services, which is part of the STFC DiRAC HPC Facility (\url{www.dirac.ac.uk}) within the RAC large project DISCSIM IV and Cambridge Service for Data Driven Discovery (CSD3).

We acknowledge the use of \textsc{Splash} \citep{SPLASH_2007} and \texttt{Sarracen} \citep{sarracen} for rendering and visualisation of figures. Some of the figures in this work make use of the colormaps in the \textit{CMasher} package \citep{cmasher}.

\section*{Data Availability}

Movies and a selection of dump files for each simulation are available at \url{https://doi.org/10.5281/zenodo.17120698}. Full \textsc{Phantom} simulation output will be made available at reasonable request to the corresponding author.



\bibliographystyle{mnras}
\bibliography{references} 


\appendix

\section{Cases without dust trapping} \label{sec:no_trapping}

\begin{figure}
    \centering
    \includegraphics{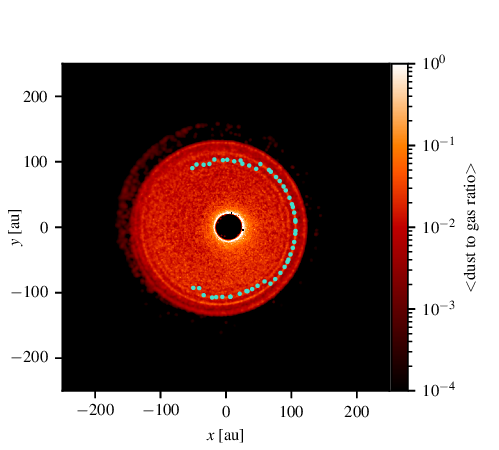}
    \caption{Column-integrated view of the dust to gas ratio at $t=1$ (16250 yr) in the \texttt{int\_retro} simulation, with the locations of the 50 tracked dust particles in the dust rings marked in blue.}
    \label{fig:ir_dust_tracked_map}
\end{figure}

\autoref{fig:ir_dust_tracked_map} shows fifty SPH dust particles in the dust rings at $t=1$ in the \texttt{int\_retro} simulation. We observe that by $t=1$ the dust substructures are beginning to dissipate and become more poorly defined. It therefore follows that these dust rings, being shorter-lived and resulting from a weaker disc-perturber interaction, will be less efficient dust traps.

\begin{figure}
    \centering
    \begin{subfigure}{\columnwidth}
        \includegraphics{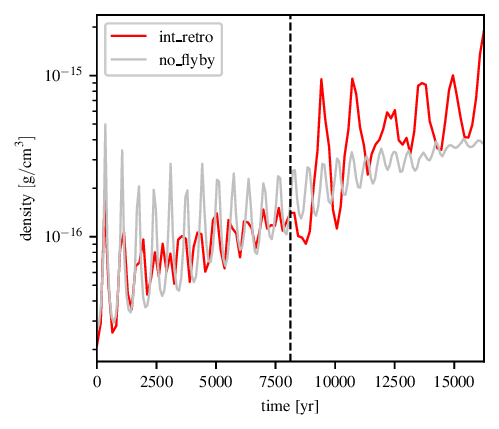}
        \caption{Dust density of tracked particles as a function of time.}
        \label{fig:ir_dust_tracked_rho}
    \end{subfigure}
    \begin{subfigure}{\columnwidth}
        \includegraphics{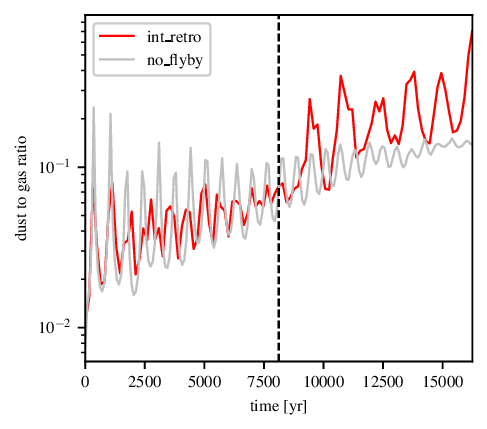}
        \caption{Dust to gas ratio of tracked particles as a function of time.}
        \label{fig:ir_dust_tracked_dtg}
    \end{subfigure}
    \caption{Mean (\subref{fig:ir_dust_tracked_rho}) dust density and (\subref{fig:ir_dust_tracked_dtg}) dust to gas ratio of 50 dust particles tracked from dust maxima in the \texttt{int\_retro} simulation, and a comparison set of 50 dust particles with initial radius $r = 125 \text{ AU}$ in the unperturbed \texttt{no\_flyby} disc (grey line). The dashed line marks $t=0.5$ for the \texttt{int\_retro} simulation.}
    \label{fig:ir_dust_tracking}
\end{figure}

\autoref{fig:ir_dust_tracking} shows the mean density and dust to gas ratio of these 50 particles over time. We see that there is no enhancement in dust density or dust to gas ratio of the tracked particles after the flyby, and furthermore that their behaviour is very similar to a comparison set of \texttt{no\_flyby} particles. We 
observe large oscillations in the dust density and dust to gas ratio of the tracked particles after the flyby, suggesting that the particles are moving in and out of dust maxima but do not remain within dust maxima for extended periods of time. As expected, the perturber's interaction with the disc in the \texttt{int\_retro} simulation is too weak to induce dust traps in the disc. 

\begin{figure}
    \centering
    \includegraphics{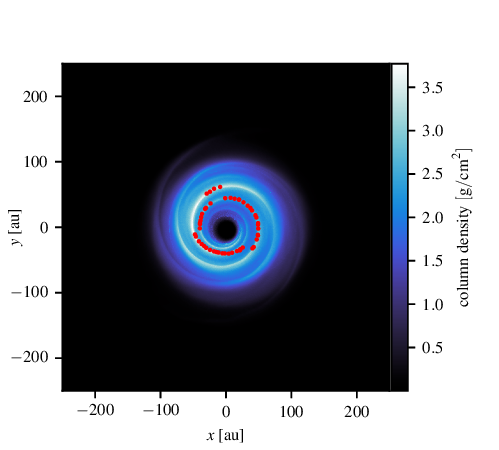}
    \caption{Column density at $t=0.6$ (15007 yr) in the \texttt{far} simulation, with the locations of the 50 tracked dust particles in the gas spirals marked in red.}
    \label{fig:far_gas_tracked_map}
\end{figure}

\begin{figure}
    \centering
    \begin{subfigure}{\columnwidth}
        \includegraphics{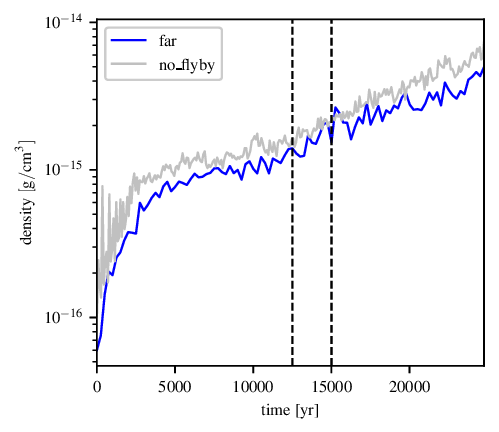}
        \caption{Dust density of tracked particles as a function of time.}
        \label{fig:far_gas_tracked_rho}
    \end{subfigure}
    \begin{subfigure}{\columnwidth}
        \includegraphics{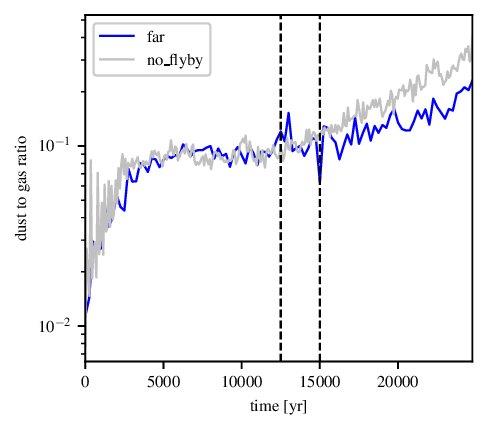}
        \caption{Dust to gas ratio of tracked particles as a function of time.}
        \label{fig:far_gas_tracked_dtg}
    \end{subfigure}
    \caption{Mean (\subref{fig:far_gas_tracked_rho}) dust density and (\subref{fig:far_gas_tracked_dtg}) dust to gas ratio of 50 dust particles tracked from gas spirals in the \texttt{far} simulation (blue line), and a comparison set of 50 dust particles with initial radius $r = 65 \text{ AU}$ in the unperturbed \texttt{no\_flyby} disc (grey line). The dashed lines mark $t=0.5$ and $t=0.6$ respectively for the \texttt{far} simulation.}
    \label{fig:far_gas_tracking}
\end{figure}

\autoref{fig:far_gas_tracked_map} shows fifty SPH dust particles in the gas spirals at $t=0.6$ in the \texttt{far} simulation. We follow the dust density and dust to gas ratio of these particles over time, as shown in \autoref{fig:pg_gas_tracking} and on their own compared to a set of particles at a similar radius in the \texttt{no\_flyby} disc in \autoref{fig:far_gas_tracking}. We see that although the particles are clearly located within the flyby-induced spiral arms, their mean dust density and dust to gas ratio are no different from those of particles in the unperturbed disc, and hence that there is \textbf{no} dust trapping in the gas spiral arms. (Note that four of the particles shown in \autoref{fig:far_gas_tracked_map} are on a disconnected spiral arm with a slightly larger radius; this does not have any meaningful effect on the average density and dust to gas ratio shown in \autoref{fig:far_gas_tracking}, because their dust density and dust to gas ratio is in fact very similar to those of the other tracked particles despite their location in a substructure at a larger radius.)

\section{Critical values for the streaming instability} \label{sec:si_more}

\begin{figure*}
    \centering
    \includegraphics{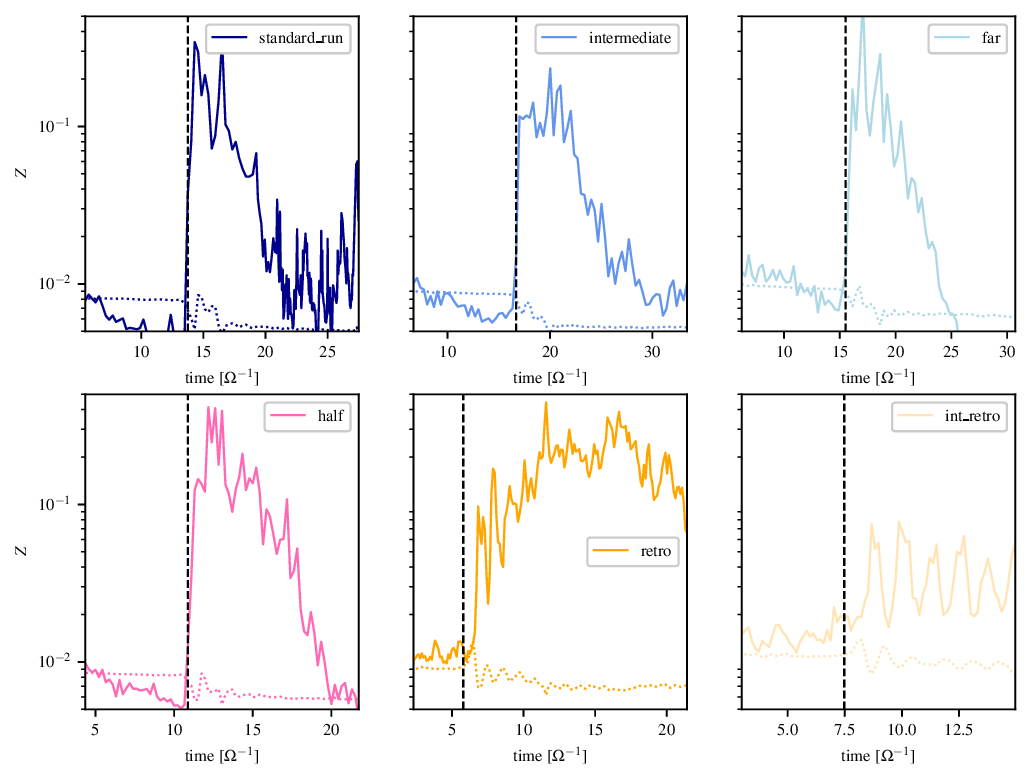}
    \caption{Solid lines: mean surface dust to gas ratio $Z$, as calculated using \autoref{eqn:surface_dtg}, of particles tracked from flyby-induced dust substructures in each simulation. Dotted lines: mean value of $Z_\text{crit}$ calculated for each particle as a function of time using \autoref{eqn:z_crit}. The dashed vertical lines mark the time of periastron for each flyby. Simulations are, from top to bottom and left to right: \texttt{standard\_run}, \texttt{intermediate}, \texttt{far}, \texttt{half}, \texttt{retro} and \texttt{int\_retro}. Note that the time axis for each simulation is scaled by the dynamical time at the location of the flyby-induced dust substructure. Note also that each plot starts a little after $t=0$ in order to give the dust particles time to settle to the midplane.}
    \label{fig:surface_dtg_grid}
\end{figure*}

\autoref{fig:surface_dtg_grid} shows the mean surface dust to gas ratio $Z$ of particles tracked from dust substructures in each flyby encounter, as well as the threshold value of $Z$ required to trigger the streaming instability as a function of time in each case, as in \autoref{fig:surface_dtg}. For the four prograde encounters we track particles from dust substructures at $t=0.6$, and for the two retrograde encounters at $t=1$, as shown in \autoref{sec:tracking_results}. (We include results from the \texttt{int\_retro} simulation for completeness, although \autoref{fig:ir_dust_tracking} confirms that there is no dust trapping in the flyby-induced dust substructures in this simulation.) 

As in \autoref{eqn:surface_dtg}, we choose to scale the time axis by the dynamical time $\Omega^{-1}$ at the location of the dust substructure, a value that is on the order of hundreds of years depending on the substructure's location. In all encounters where dust trapping is seen, the surface dust to gas ratio of the tracked particles exceeds the threshold value required to trigger the SI for at least $10\Omega^{-1}$. From this we can conclude that flybys with a range of periastron distances, perturber masses and orientations are all capable of triggering the SI in the perturbed disc.

\bsp	
\label{lastpage}
\end{document}